\pdfoutput=1 

\documentclass[aip, reprint, floats, floatfix]{revtex4-1}

\usepackage{graphicx}	
\usepackage{bm}		

\usepackage{natbib}

\usepackage{color}
\usepackage[normalem]{ulem}		

\begin{document}

\title[Global communication pathways of the human brain]
{The global communication pathways of the human brain transcend the cortical - subcortical - cerebellar
division}

\author{Julian Schulte}
\affiliation{Center for Brain and Cognition, Pompeu Fabra University, Barcelona, Spain.}
\affiliation{Department of Information and Communication Technologies, Pompeu Fabra University, Barcelona, Spain.} 

\author{Mario Senden}
\affiliation{Department of Cognitive Neuroscience, Faculty of Psychology and Neuroscience, University of
Maastricht, Maastricht, The Netherlands.}
\affiliation{Maastricht Brain Imaging Centre, Faculty of Psychology and Neuroscience, Maastricht University,
Maastricht, The Netherlands.}

\author{Gustavo Deco}
\affiliation{Center for Brain and Cognition, Pompeu Fabra University, Barcelona, Spain.}
\affiliation{Department of Information and Communication Technologies, Pompeu Fabra University, Barcelona, Spain.} 
\affiliation{Instituci\'o Catalana de la Recerca i Estudis Avan\c{c}ats (ICREA), Barcelona, Spain.} 

\author{Xenia Kobeleva}
\affiliation{Computational Neurology Group, Department of Neurostimulation, Ruhr University Bochum, Bochum,
Germany.}

\author{Gorka Zamora-L\'opez}  \email{gorka@Zamora-Lopez.xyz}
\affiliation{Center for Brain and Cognition, Pompeu Fabra University, Barcelona, Spain.}
\affiliation{Department of Information and Communication Technologies, Pompeu Fabra University, Barcelona, Spain.} 
\affiliation{Department of Complex Systems, Institute of Computer Science of the Czech Academy of Sciences, Prague, Czech Republic.}


\begin{abstract} 	
Understanding how cortex, subcortex and cerebellum integrate is a major challenge for neuroscience, however, studies of the brain's structural connectivity have mostly focused on cortico-cortical links. Here, we used diffusion imaging to construct the structural connectome of the entire human brain including 360 cortical, 233 subcortical, and 125 cerebellar regions of interest (ROIs). We found that the brain forms a modular and hierarchical network architecture, organized into modules of mixed cortical, subcortical and/or cerebellar regions, and whose cross-modular pathways are centralized through highly connected hub ROIs (a `rich-club'). This global rich-club is subcortically dominated and, surprisingly, composed of hub ROIs from all subcortical structures rather than one region like the thalamus, centralizing the communication pathways. This study improves our understanding of the human brain's organization. It provides structural evidence to question the prevalent cortico-centric notion by revealing a connectome centered at the subcortex but made of transversal pathways.
\end{abstract}


\maketitle

\newpage
\section*{Introduction}

Complex cognition emerges from the coordinated activity of distributed brain regions. Historically, a dominant cortico-centric perspective of the brain largely confined higher functions to the neocortex, while subcortical and cerebellar structures were considered to play supportive roles~\cite{Parvizi_Corticocentric_2009,Pessoa_Understanding_2014}. Converging evidence now critically challenges this conception by demonstrating, for example, the crucial contribution of the basal ganglia to cognition~\cite{Obeso_BasalGanglia_2014}, of the hippocampus to decision-making and reward~\cite{Attaallah_Hippocampus_2024}, and of the cerebellum to cognitive operations~\cite{Palesi_Pathways_2015, Palesi_Contralateral_2017}. An important missing gap to fully embrace this emerging holistic view of brain function is to understand how the cortex, subcortex and cerebellum communicate with each other at the systems level.

The study of structural connectivity derived from white-matter fibers has revealed that the brain is shaped into a hierarchically modular network, centralized by a set of hub regions termed the `rich-club'~\cite{Zamora_Hubs_2010, Zamora_FrontReview_2011, Heuvel_HubsHuman_2011}. The functional relevance of these hubs has been empirically corroborated for multi-sensory and cognitive integration~\cite{Senden_RichClub_2014, Senden_RichClub_2017, Bertolero_Modular_2015}, and their dysfunction has been associated with various neuropsychiatric diseases~\cite{Crossley_Hubs_2014}. So far, these connectivity studies have excluded the cerebellum and, when included, the subcortical structures were treated as a single node of the network introducing an unbalanced mixture of resolutions. For example, the thalamus broadly connects with the cortex but these projections originate from distinct subthalamic nuclei~\cite{jones2001thalamic, Kumar_Thalamic_2023}. Therefore, considering the whole thalamus as a single node overestimates its network relevance and affects the outcome of computational models. On a local scale, detailed descriptions of cortico-subcortical links have been reported using either tractography or functional connectivity, but usually focusing on the connectivity of one structure alone, such as the thalamus~\cite{Kumar_Thalamic_2023, Bell_Subcortical_2016}, basal ganglia~\cite{Bell_Subcortical_2016} or brainstem~\cite{Hansen_Brainstem_2024}. Accordingly, while the central role of subcortical structures has been highlighted, no exhaustive investigation has yet been undertaken to describe how the various subcortical structures are interconnected at the macroscopic level nor to clarify the place each takes in the integrated architecture of the entire brain.

Here, we employed diffusion imaging and tractography to investigate the structural architecture of the human brain encompassing the cortex, the cerebellum and nine subcortical structures. With the brain parcellated into 718 regions of interest (ROIs), this network presented sufficient resolution to separate subcortical structures into multiple ROIs allowing to explore their detailed connectivity. Our analyses revealed that this brain-wide human connectome follows a modular and hierarchical architecture mirroring the one reported for cortico-cortical connectivity alone~\cite{Scannell1993, Scannell1995, Hilgetag_Clusters_2000, Hilgetag_Clustered_2004, Zamora_Hubs_2010, Zamora_FrontReview_2011, Heuvel_HubsHuman_2011, Betzel_Modular_2017, Suarez_Taxonomy_2022, Puxeddu_Relation_2024}, but with significant differences. First, the human brain is organized into macroscopic network modules, composed of intermixed cortical, subcortical and/or cerebellar ROIs thus transcending their individual boundaries. Second, the pathways between these modules are centralized through a communication core (a `rich-club') that contains cortical and subcortical regions but, importantly, is dominated by subcortical rather than by cortical ROIs. In fact, we showed that \emph{in-silico} lesions of subcortical hubs greatly disrupted the communication throughout the network, significantly more than the lesion of cortical hubs. Last, and surprisingly, we found subcortical structures to be internally heterogeneous. Their ROIs show a diversity of local and global connectivity profiles indicating they play differential functional roles. A relevant consequence of this internal heterogeneity is that all subcortical structures contribute---partly---to the global rich-club, in contrast to previous reports prone to place one structure (e.g., thalamus or basal ganglia) at the center of the brain-wide communication. Altogether, our study improves the current understanding of the human brain's organization. It introduces a novel and complementary view, providing structural arguments to dispute the prevalent cortico-centric notion that regards subcortical structures and cerebellum as \emph{mere} supporters of cortical function. Instead, our results reveal a network architecture centered at the subcortex but made of transversal pathways.

\section*{Results}
Structural connectivity data was obtained via diffusion-weighted imaging for 32 healthy human participants from the Human Connectome Project (HCP)~\cite{VanEssen_Development_2018} and parcellated into 718 regions of interest (ROIs). Of these, 360 correspond to cortical, 233 to subcortical, and 125 to cerebellar cortex (see~\citet{Ji_Mapping_2019} for details). This parcellation provides the internal division of subcortical structures into several ROIs such as the bilateral nucleus accumbens (n=13), brainstem (n=47), caudate nuclei (n=17), diencephalon (n=40), hippocampus (n=29), globus pallidus (n=20), putamen (n=18), thalamus (n=38), and amygdala (n=11). Tractography data was filtered in order to exclude false-positive tracts related to cerebellar connections~\cite{Ramnani_CorticoCereb_2006, Palesi_Pathways_2015, Palesi_Contralateral_2017, DAngelo_Physiology_2018}. A population averaged, binary structural adjacency matrix was estimated employing a distance-aware consensus algorithm. For the rest of the paper, we investigate the properties of this population-level connectivity.

\subsection*{Network division of the brain-wide structural connectivity}

\begin{figure*}[ht]
	\centering
	\includegraphics[width=1.0\textwidth,clip=]{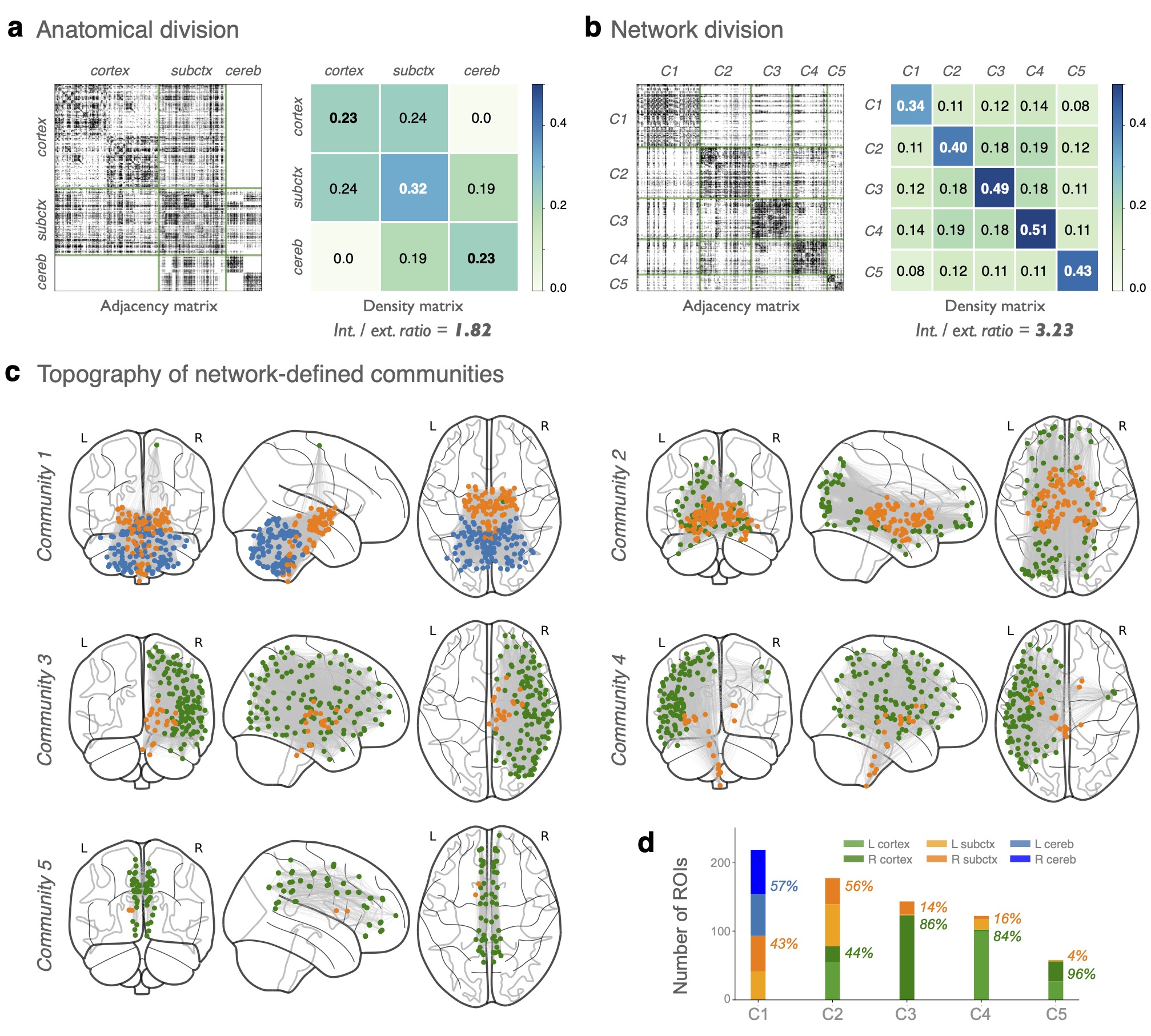}
 	\caption{		\label{fig:Figure1}
	{\bf Modular organization of the brain-wide connectome.}
	Structural connectivity matrix and corresponding intra-/inter-modular connection densities for the brain-wide connectivity sorted according 
	{\bf a}, to the traditional anatomical division of the brain into cortex, subcortex and cerebellum, and 
	{\bf b}, to the optimal network division into five communities found by community detection algorithms. The optimal network division shows larger intra-modular densities and lower cross-modular connection probabilities than for the anatomical division. 
	{\bf c}, Topographical distribution of the five network communities shows that communities 1-4 vertically stretch over cortical, subcortical, and cerebellar boundaries; communities 3 and 4 are lateralized comprising cortical and subcortical regions and the 5th is made of medially arranged cortical ROIs.
	{\bf d}, Quantitative composition of the five network communities in terms of cortical, subcortical or cerebellar ROIs, with hemispheric information added to highlight the lateralization of communities 3 and 4.
	} 
\end{figure*}

We begin by inquiring whether the anatomical division of the brain into cortex, subcortex and cerebellum is also reflected in the organization of its structural connectivity. For that, we estimated the average probability that an ROI connects to other ROIs in the same anatomical category (diagonal entries of density matrix in Fig.~\ref{fig:Figure1}a) and to ROIs of other categories (extradiagonal entries). In terms of network architecture, a good division requires intra-category probabilities to be notably larger than inter-category probabilities. However, we see that cortical ROIs are equally likely connected with each other as with subcortical ROIs (cor-cor: 0.23; cor-sub: 0.24), while cerebellar ROIs are very likely connected to subcortical ones too (cer-cer: 0.23; cer-sub: 0.19). This observation indicates that this traditional, anatomical classification does not represent a faithful organization of the brain's connectivity into network modules.

To identify a more appropriate description of the brain's systems level organization, we apply a community detection algorithm (Leidenalg~\cite{Traag_Leidenalg_2019}) to automatically cluster the 718 ROIs based on network properties alone. This algorithm aims at grouping together nodes that are more densely connected with each other than with the nodes in other groups, by maximizing an intra-/inter-connectivity cost function (e.g., Newman modularity). The optimal solution (resolution parameter $\gamma = 1.0$, $Q = 0.232$; see Supplementary Fig.~\ref{fig:FigureS1}) returned a division of the brain-wide connectome into five communities (modules or clusters). The adjacency matrix is shown in Fig.~\ref{fig:Figure1}b, reorganized for this optimal partition alongside its corresponding density matrix. This network-based division clusters the brain notably better than the anatomical division (Fig.~\ref{fig:Figure1}a). The diagonal entries now display larger internal connection probabilities (0.34 -- 0.51) and lower cross-modular connection probabilities (0.08 -- 0.19). On average, under this network-based division, ROIs are 3.23 times more likely connected to other ROIs in their community than with ROIs in other communities.

The five network communities consist of a mixture of regions surpassing the cortical, subcortical, and cerebellar boundaries (Figs.~\ref{fig:Figure1}c,d). Community C1 contains left and right, ventrally located subcortical (43\%) and cerebellar ROIs (57\%). The entire cerebellum is embedded into C1. Community C2 comprises cortical (44\%) and subcortical (56\%) ROIs on both hemispheres. Notably, the cortical regions participating in C2 are located in frontal and posterior lobes, Fig.~\ref{fig:Figure1}c. Communities C3 and C4 display strong lateralization (right and left) and are predominantly composed of cortical ROIs (84\% and 86\%), with fewer subcortical ROIs (16\% and 14\%). Finally, a smaller community C5 is medially located comprising left and right cortical ROIs (96\%) and a few subcortical ROIs (4\%). These results provide evidence that, from a network perspective, the brain-wide connectome is organized into network modules that transcend the cortical, subcortical, and cerebellar boundaries.

\subsection*{Distribution of subcortical nuclei across network communities}

\begin{figure*}[ht]
	\centering
	\includegraphics[width=1.0\textwidth,clip=]{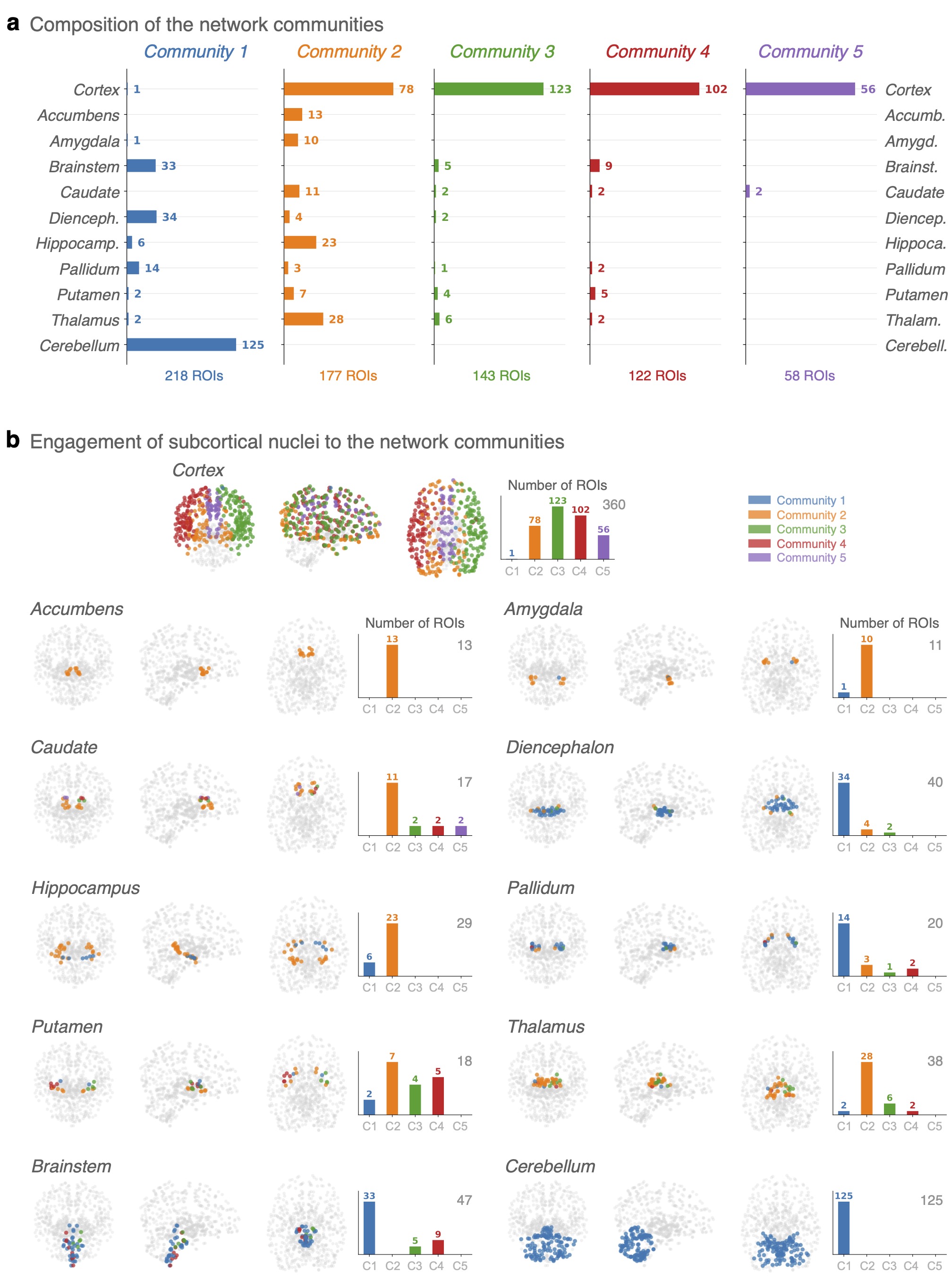}
\end{figure*}
\begin{figure*}
 	\caption{		\label{fig:Figure2}
	{\bf Relation between anatomical and optimal network communities.}
	{\bf a}, Composition of the five optimal network communities in terms of anatomical components, including the division of the subcortex into nine nuclei. Bars indicate the number of ROIs from each anatomical component that was classified into the corresponding network community. Numbers at the bottom summarize the size of the network communities, in total number of ROIs. 
	{\bf b}, Internal division of the nine subcortical structures, the cortex and the cerebellum in terms of the five network-based communities. Topographical maps show that---in most cases except for Accumbens and cerebellum---the ROIs of an
anatomical component usually fall into different communities, evidencing an internal distribution of anatomical structures and their participation into different circuits. Bars summarize the number of ROIs that each anatomical component devotes to distinct network communities (C1~--~C5). Gray numbers indicate the size (in number of ROIs) of the anatomical components.	} 
\end{figure*}

We now take a closer look at the composition of these five network communities to better identify the role of subcortical structures in this division, Fig.~\ref{fig:Figure2}a. C1 is mainly made of ROIs from the cerebellum, brainstem and diencephalon, followed by contributions from pallidum and hippocampus. Community C2 is made of frontal and posterior cortical ROIs, and a mixture of subcortical nuclei: mainly thalamus, hippocampus, accumbens, caudate and amygdala, and to a lesser extent, also the putamen, diencephalon and pallidum. Communities C3 and C4 are made of lateralized cortical regions together with subcortical ROIs from brainstem, thalamus, putamen, pallidum and caudate. Last, C5 is practically a cortical community.

Given that the subcortical structures are smaller than cortex and cerebellum (in number of ROIs), a relative interpretation of these results is needed. Thus, Fig.~\ref{fig:Figure2}b shows how the ROIs of each structure are devoted to the different communities. The putamen is the most segregated subcortical structure with its 18 ROIs very much distributed among C1 -- C4. On the opposite side, the accumbens and the amygdala contribute (almost) all their ROIs to C2. The remaining structures show an intermediate situation, with most of their ROIs assigned to one community and fewer ROIs contributing to other communities.

The involvement of subcortical structures in distinct network communities is relevant because it implies that despite their small size, subcortical structures are internally diverse, and indicates their participation in many functional circuits.

\subsection*{Contribution of individual ROIs to cross-modular connectivity}

\begin{figure*}[ht]
	\centering
	\includegraphics[width=1.0\textwidth,clip=]{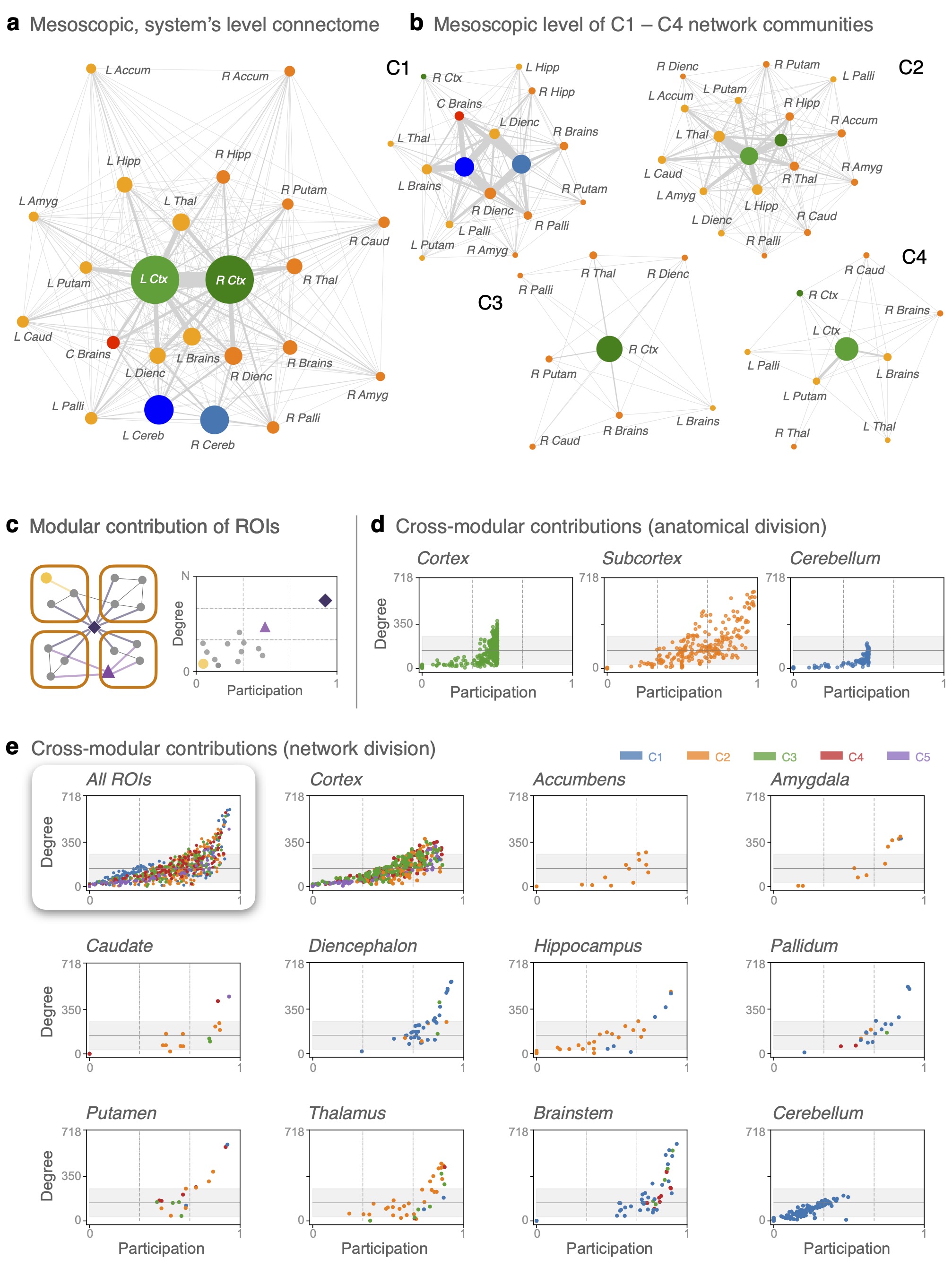}
\end{figure*}
\begin{figure*}
 	\caption{		\label{fig:Figure3}
	{\bf Contribution of anatomical components to cross-modular connectivity.}
	{\bf a, b}, Visualization of the structural connectivity at the level of anatomical components comprising the cortex, the cerebellum and nine subcortical structures; {\bf a}, for the brain-wide network and {\bf b}, the subnetworks representing network communities C1 to C4. 
	Community C5 is omitted for its simplicity as it only contains three regions. Node diameters are proportional to the number of ROIs to each anatomical component.
	 Link widths reflect the number of ROIs connected within the two components. 
	{\bf c}, Illustration of different roles that nodes can play within a modular and hierarchical network, and their characterization. Nodes can be either locally connected inside a module (yellow dot), be well rooted in a community but share connections to other communities (purple triangle), or uniformly participate over all communities (dark purple diamond). 
	These differential contributions are typically characterized by two parameters allowing to project each node into a 2D map (axes on the right panel). (i) The degree, informing of the overall centrality of a node and (ii) the participation index, indicating how a node spreads its links among communities. Participation 0.0 means the node is only locally connected in one module (e.g., the blue dot) while participation 1.0 means a node is equally likely connected to all communities (e.g., the purple diamond). Locally connected nodes are expected to play more specialized functions while hubs with large participation are usually associated with multifunctional roles due to their access to a wider range of information. 
	{\bf d}, Mapping of the roles of all the ROIs, considering the network is divided into cortex, subcortex and cerebellum. For visual clarity, the 718 ROIs are presented in three separate maps for cortical, subcortical and cerebellar regions. The solid horizontal lines represent the average node degree and gray shadows the range for one standard deviation. 
	{\bf e}, Mapping of the roles for all the ROIs, considering the network is divided into the five network communities. Top left panel shows all the 718 ROIs on a single map. The remaining panels show the same results but individually for the ROIs of each of the 11 anatomical components studied. 
	ROIs are colored according to the network community they were assigned to.
	} 
\end{figure*}

So far, we have explored the overall division of the brain into network communities and their composition. Now, we investigate how these brain-wide communities connect with each other. Network visualization algorithms allow for an approximate representation of a network's architecture by arranging their nodes according to their relative importance and mutual similarity. The whole-brain connectome is displayed in Fig.~\ref{fig:Figure3}a (Force Atlas 2 algorithm~\cite{Jacomy_ForceAtlas_2014}). In this visualization the network appears centered around the two cortical hemispheres (in green) largely connected to the thalamus and the putamen on the one hand, with the brainstem, diencephalon and pallidum mediating the cortico-cerebellar pathways on the other hand. The smaller subcortical structures tend to be placed around the periphery, with those contributing mostly to the C2 community (e.g., accumbens, amygdala, hippocampus and thalamus; see Fig.~\ref{fig:Figure2}b) occupying the upper space while the structures more associated to C1 (e.g., brainstem, diencephalon and pallidum; Fig.~\ref{fig:Figure2}b) are placed towards the bottom, near the cerebellum.

This spatial separation is reflected in more detail when observing the subnetworks formed individually by the network communities C1 and C2, Fig.~\ref{fig:Figure3}b. The core of C1 exhibits substantial interconnectivity between the cerebellum, the brainstem (specially its central part) and the diencephalon. Instead, C2 is governed by strong interconnections between the two cortical hemispheres and a substantial number of connections to the hippocampus, thalamus and caudate. C3 and C4 form star-like structures centered on either right or left cortical ROIs.

Although these layouts are informative, they are affected by the size differences (in number of ROIs) of the anatomical components. Looking at connection probabilities instead of total link numbers, some smaller areas like the putamen, amygdala, and thalamus become more central, highlighting their key role in connecting different parts of the brain (see Supplementary Fig.~\ref{fig:FigureS2}). 

We now clarify in more detail the position that every ROIs takes within the brain-wide connectome. For that, we map the 718 ROIs based on two parameters~\cite{Guimera_RolesJSM_2005, Zamora_FrontReview_2011, Klimm_Roles_2014}: their degree (i.e., hubness) and their participation index which quantifies how (un-)evenly the connections of an ROI are distributed across modules. See Fig.~\ref{fig:Figure3}c for a schematic description. Hence, the participation index of a node depends on how the network was divided into modules in the first place.

Considering the anatomical division (Fig.~\ref{fig:Figure3}d), cortical and cerebellar ROIs tend to occupy the lower-right and the central sectors of the map, corresponding to nodes with local or intermediate cross-modular connections. The reason is the absence of direct connections between cerebellum and cortex. Thus, under this anatomical view, cortical and cerebellar ROIs span their connections, at most, among two of the three modules (i.e., either cortex and subcortex, or cerebellum and subcortex) and thus attain a maximal participation index of 0.5. On the other hand, several subcortical ROIs occupy the upper-right sector of the map which is characteristic of supramodal hubs (e.g., purple diamond in Fig.~\ref{fig:Figure3}c). Importantly, this also indicates that the brain-wide communication is mainly sustained by subcortical regions.

Considering the network-based division (Fig.~\ref{fig:Figure3}e), the 718 ROIs distribute along the diagonal axis of the 2D map (top-left panel), which is a signature of a modular and hierarchical organization~\cite{Zamora_FrontReview_2011, Klimm_Roles_2014, Zamora_FComplexity_2016}. Disentangling the results individually by anatomical structures, the ROIs of the cortex display a wider variety of roles as compared to the results under the anatomical division. It now includes some hubs of high participation (up to 0.86). On the contrary, cerebellar ROIs still fall into the lower-left sectors of the map, as in fact, cerebellar ROIs span their connections mostly among ROIs in C1 and C2. For the subcortical structures, the results show a rich variety of connectivity roles. For example, the accumbens and the amygdala were entirely classified into community C2, however, some of their ROIs display both large degree and high participation, indicating their connections span throughout all communities and thus taking a central role in the network. The caudate and the putamen are two other small nuclei that contain ROIs with high degree and participation, although in this case it was more expected given that their ROIs were distributed among C2 -- C5 and C1 -- C4 respectively, Fig.~\ref{fig:Figure2}.

Altogether, these results show that subcortical structures are internally heterogeneous. Their ROIs occupy different positions in the degree-participation maps, thus indicating they play differential roles in the network architecture. Some ROIs form mostly local connections (characterized by low degree and participation) and others span their connections across multiple communities (large degree and participation) becoming crucial for the centralization of the brain-wide pathways.

\subsection*{Subcortical hubs dominate the communication core of the brain-wide connectome}

\begin{figure*}[ht]
	\centering
	\includegraphics[width=1.0\textwidth,clip=]{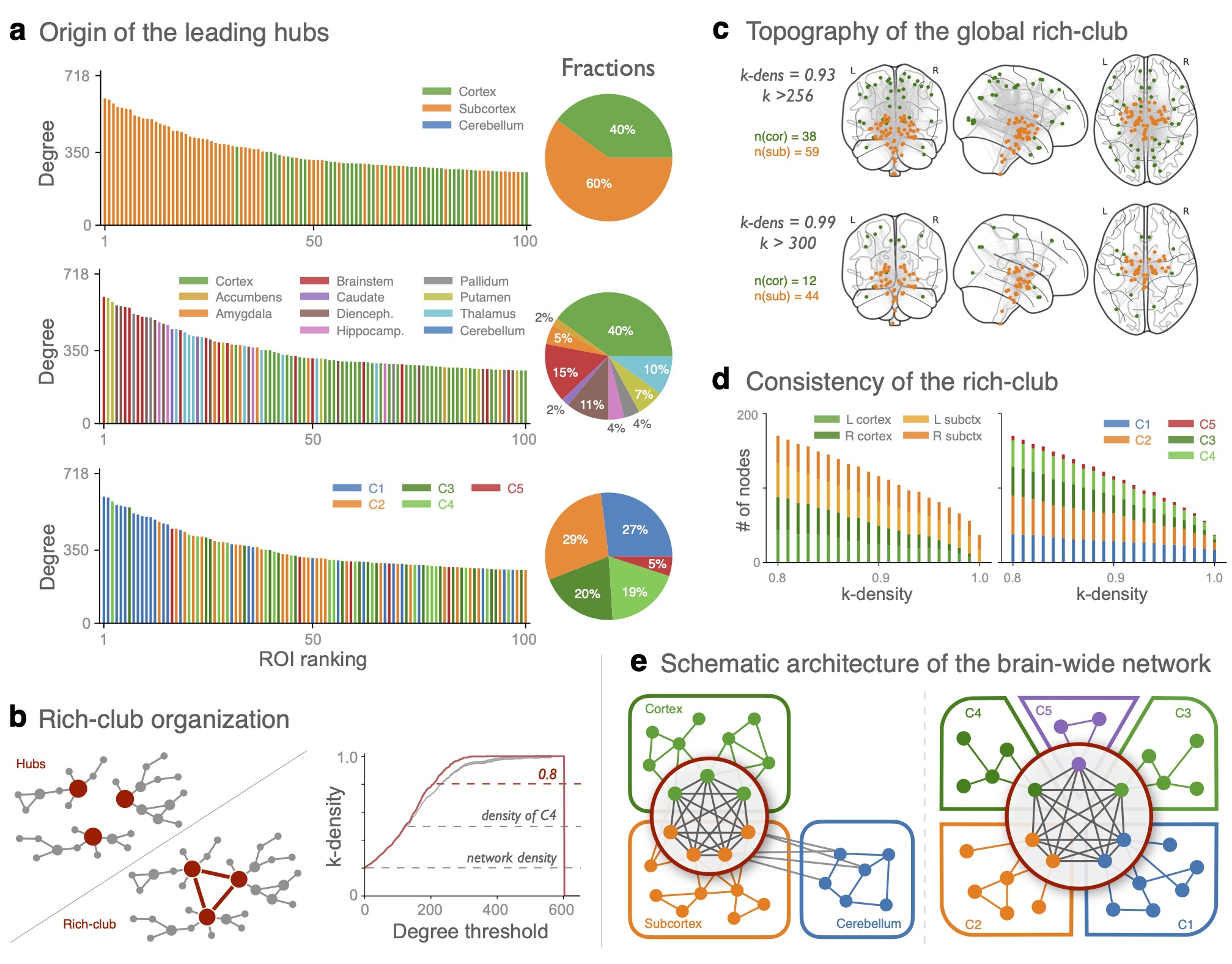}
 	\caption{		\label{fig:Figure4}
	{\bf Hierarchical centralization of cross-modular connectivity.}
	{\bf a}, Analysis of the 100 lead hub ROIs of the brain-wide network. Bar panels display their degree---sorted in decreasing order---with the bars colored according to the community to which the ROI belongs, in the three different divisions. Pie charts show the fraction of leading hubs among the subsequent communities. 
	{\bf b}, Evidence for a rich-club supra-organization among the leading hubs. Left panel, conceptual illustration of a rich-club, formed when the leading hubs are densely connected with each other. Right panel, evolution of the link density for the subnetworks formed by ROIs with degree larger than $k$ (nodes with degree smaller than $k$ are removed), for all possible thresholds $k = 0, 1, \ldots, 718$. Red curve is the evolution for the brain-wide network, gray curve is for degree-conserving randomized samples (100 independent realizations each point). The threshold for $k = 0$ represents the entire network, thus, the curves begin from a density of 0.2. Their monotonic increase up to a density of 1.0 evidences that the leading hubs eventually form an all-to-all connected core, centralizing the communication pathways. 
	{\bf c}, Topographical location of the rich-club hubs and their mutual links, for two different density thresholds: at $k > 256$ (subgraph density of 0.93) and $k > 300$ (subgraph density of 0.99). Cortical ROIs are shown as green dots and subcortical ROIs in orange. 
	{\bf d}, The composition of the rich-club at different levels of internal density, from 0.8 to 1.0 (degree thresholds, $k \geq 208$ to $k = 365$), displayed for the anatomical (left or right cortex, subcortex, cerebellum) and the network divisions (C1~--~C5). Bars indicate the number of ROI hubs forming the rich-club at each density level, and how many of them belong to the different anatomical or network communities. 
	{\bf e}, Schematic representation of the modular and hierarchical organization of the brain-wide structural connectivity, in the light of either the anatomical or the optimal network divisions.
	} 
\end{figure*}

The results of Fig.~\ref{fig:Figure3} highlight the presence of hub ROIs in the cortex and all subcortical structures. We now examine how these hubs help elucidate the brain-wide network architecture. Ranking the 100 most connected ROIs together, Fig.~\ref{fig:Figure4}a top panel, reveals that the leading hubs are mainly subcortical (60\%), followed by cortical hubs (40\%). Remarkably, no cerebellar ROI is present among the top global hubs. Distinguishing these leading hubs by subcortical structure (middle panel) confirms that all subcortical structures contain hub ROIs (regardless of their size) and thus, no single structure can be considered responsible for the centralization of the brain-wide connectivity. From the network-based division (bottom panel) the leading hubs are broadly distributed with the C1 -- C4 communities containing 19\% -- 29\% of the hubs each.

Additionally, we find clear evidence that these leading hubs form a rich-club supra-community, see Fig.~\ref{fig:Figure4}b (left panel) for an illustration. By iteratively removing the ROIs with less connections, the internal density of the remaining subnetwork (made of the ROIs with larger degree) rapidly increases, Fig.~\ref{fig:Figure4}b (right panel). There are 170 ROIs with degree $k > 208$, forming a core of density 0.8. In comparison, this is 56\% denser than C4, which is the densest (0.51) of the five network communities. The 100 leading hubs displayed in Fig.~\ref{fig:Figure4}a form a rich-club with internal density of 0.93; their topographical location is shown in Fig.~\ref{fig:Figure4}c.

As the threshold is increased, the number of hubs in the rich-club decreases, Fig.~\ref{fig:Figure4}d. This happens mainly at the expense of cortical hubs (left panel), emphasizing the central role of the subcortex in the brain-wide connectome. However, the representation of network communities in the rich-club remains balanced across thresholds (right panel).

The analysis performed here shows that, at the systems level, the connectivity of the entire brain follows similar organization principles as those observed for the cortico-cortical connectivity alone~\cite{Zamora_Hubs_2010, Zamora_FrontReview_2011, Heuvel_HubsHuman_2011}: it is structured into a handful of network modules with the cross-modular connectivity centralized via hub ROIs, which form a spatially delocalized supramodule at the core, Fig.~\ref{fig:Figure4}e. However, in this global case, we found that it is the subcortical ROIs---instead of the cortical areas---which dominate the communication core of the network, while the cerebellum is excluded from the rich-club, Fig.~\ref{fig:Figure4}e (left panel). Particularly relevant is the fact that no single subcortical structure centralizes the global pathways, but instead, subcortical hubs appear distributed across the different structures. This is further highlighted from the viewpoint of the network-based division, with the rich-club hubs evenly dispersed among the network communities.

\subsection*{Global integration and segregation in the brain-wide connectome}

\begin{figure*}[ht]
	\centering
	\includegraphics[width=1.0\textwidth,clip=]{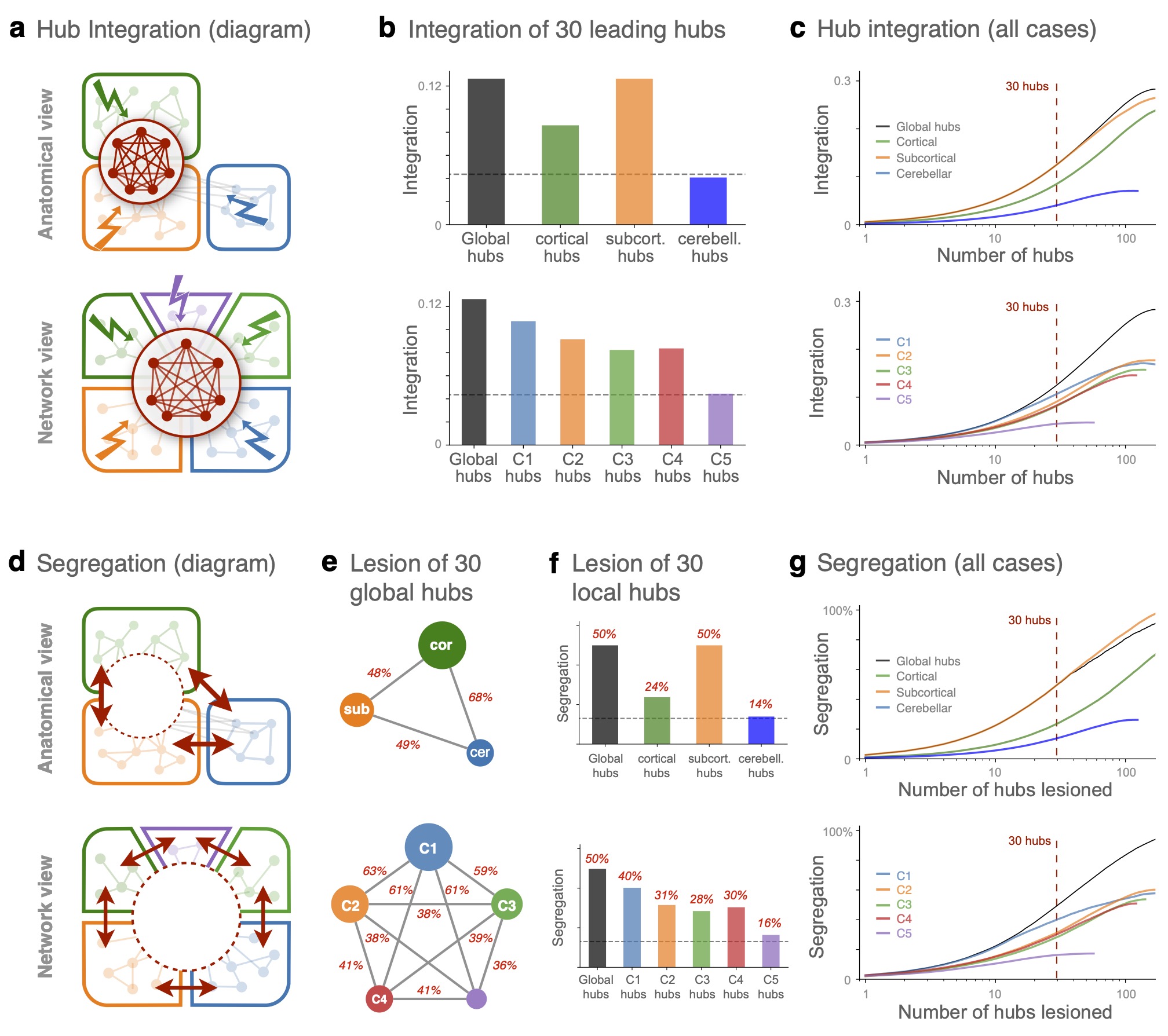}
 	\caption{		\label{fig:Figure5}
	{\bf Integrative and segregative consequences of the brain-wide network architecture.}
	{\bf a}, Schematic illustration of the integration capacity metric for a set of ROIs, in this case, the network hubs. Illustration for both the anatomical or the network divisions. Integration is defined as the joint response of a group of ROIs (the hubs), to stimuli applied in all other ROIs. 
	{\bf b}, Integration capacity of the 30 leading hubs, chosen from the global ranking (black bars) or chosen from the local ranking at each community (colored bars). Top, for the anatomical division into cortex, subcortex, and cerebellum; bottom, for the leading hubs within the network communities C1~--~C5. Dashed lines indicate the average integration capacity of 30 randomly chosen ROIs, regardless of their hubness (average of 100 realizations). 
	{\bf c}, Integration capacity of the $n$ leading hubs, with $n$ from 1 to 170. Black lines, choosing hubs from their global ranking and, colored lines, choosing the hubs of a given community, see Fig.~\ref{fig:Figure4}a (top and bottom). Notice that the sizes of cerebellum, C3, C4 and C5 are smaller than 170, thus, their last points reflect the integration capacity of the entire community. 
	{\bf d}, Schematic representation of the modular segregation metric, in both the anatomical and the network divisions. Segregation is defined as the conditional loss of communication between modules, given that a set of $n$ ROIs has been lesioned. In this case, a set of hubs is lesioned. 
	{\bf e}, Loss of communication between the modules after the leading 30 global hubs have been lesioned. Numbers indicate the loss between each pair of modules. 
	{\bf f}, Segregation suffered by the network after lesion of the 30 leading hubs, selected from the global ranking (black bars) or locally selected within a community (colored bars). Dashed lines indicate the average modular segregation after 30 randomly chosen ROIs are lesioned (average of 100 realizations). 
	{\bf g}, Modular segregation suffered by the network after the n leading hubs have been lesioned, with $n$ from 1 to 170. Black lines, choosing hubs from their global ranking and, colored lines, choosing them within each community. Notice that the sizes of cerebellum, C3, C4 and C5 are smaller than 170, thus, their last points reflect the segregation of the network after the entire community has been lesioned.
	} 
\end{figure*}

We conclude this study by exploring functional implications of the brain-wide architecture uncovered and depicted in Fig.~\ref{fig:Figure4}e. For that, we examine its capacity to host integrative and segregative processes, and its vulnerability to targeted lesions. We evaluate integration and segregation following an in-silico perturbative approach that assesses network responses to external stimuli~\cite{Zamora_Hubs_2010}. This approach accounts for how brain regions influence each other through all possible pathways, beyond one-to-one interactions~\cite{Gilson_DynCom_2019, Zamora_Chaos_2024}.

Integration is defined as the joint response of a set of hubs to stimuli applied at the rest of the network, see Fig.~\ref{fig:Figure5}a for an illustration. Considering the thirty leading hubs, either globally or locally per module, the integration capacity of subcortical hubs is clearly larger than that of cortical and cerebellar hubs, Fig.~\ref{fig:Figure5}b (top). We estimate the integration capacity for different hub set sizes by sequentially adding one hub at a time, Fig.~\ref{fig:Figure5}c. The integration capacity of subcortical hubs (top panel, orange curve) is the largest across all hub set sizes, practically explaining the integration capacity of the global hubs (black curve), followed by that of cortical hubs (green curve). The integration capacity of the cerebellar hubs (blue) is very small in all cases. From the viewpoint of the network division, the integration capacity of hubs (selected by community) is more evenly distributed, Figs. 5b,c (bottom panels). Although C1 displays larger integration than the rest, the curves for integration capacity of C1 -- C4 communities remain close to each other at all hub set sizes considered, Fig.~\ref{fig:Figure5}c (bottom).

We define segregation as the loss of influence between communities, after a number of hubs has been lesioned~\cite{Zamora_Hubs_2010}, see Fig.~\ref{fig:Figure5}d for illustration. For example, the influence of C1 on C2 is first measured in the original (healthy) network. Then, a set of hubs is lesioned and the influence is re-evaluated. Segregation is determined as the relative difference (loss) of influence between C1 and C2, from healthy to lesioned. When the thirty leading global hubs are lesioned, the segregation suffered by the network is drastic: 48 -- 68\% of influence is lost between the three anatomical communities and 38 -- 63\% among the network-based ones, Fig.~\ref{fig:Figure5}e. The largest loss occurs between cortex and cerebellum (68\%). On the same order (63\%) is the loss of communication incurred between C1 and C2 network communities. The total segregation incurred after lesion of the 30 leading global hubs is 50\%, Fig.~\ref{fig:Figure5}f (black bars). This level of segregation is striking taking into account that thirty hubs represent only 4\% of all the ROIs. Selecting the hubs per community, the largest segregation (50\%) occurs when subcortical hubs are lesioned (top panel, orange bar), which is twice the segregation incurred by the lesion of cortical hubs (24\%, green bar). Selecting hubs following network communities, Fig.~\ref{fig:Figure5}f (bottom), again we identify a more even segregation after targeted C1 -- C4 hub lesions, as for their integration capacity.

Cumulative lesion of ROIs, from largest to smallest degree, confirms that subcortical lesions are the most disruptive for the communication in the network, Fig.~\ref{fig:Figure5}f (top). For sizes up to 100 ROIs, the lesion of subcortical hubs causes 60--140\% larger segregation than the lesion of cortical hubs alone. However, from the viewpoint of the network-based communities (bottom), local lesions to ROIs within the C1 -- C4 communities lead to similar levels of segregation.

These results confirm the central importance of subcortical regions to sustain brain-wide communication. Also, they show that the modular and hierarchical network organization defined by the connectivity (Fig.~\ref{fig:Figure4}e, right) constitutes a more balanced distribution of the brain into large-scale circuits.

\section*{Discussion}

A major challenge for neuroscience today is to better understand how cortex, subcortex and cerebellum work together. This anatomical division is an essential notion of the brain's organization, evidenced by their differentiated cytoarchitecture, cellular composition and evolution~\cite{pinel2018biopsychology}. However, there is more to the brain's structure than meets the eye, hidden in the complex manner that white matter shapes the pathways between distant regions, and only discernible through suitable network analyses. Here, we employed high-quality diffusion imaging and tractography from the Human Connectome Project, and a parcellation in 718 regions of interest (ROIs)~\cite{Ji_Mapping_2019}, in order to uncover the macroscopic connectional architecture of the entire brain.

\subsection*{Intermixed modular organization}

Subcortical regions are known to heavily interlink with the cortex via a multitude of pathways involving the amygdala~\cite{Phelps_Amygdala_2005}, the hippocampus~\cite{Maller_Hippocampal_2019, Huang_Extensive_2021}, the basal ganglia~\cite{Alexander_Basal_1986, Haber_Basal_2009}, the superior colliculus~\cite{Benavidez_Superior_2021}, and the thalamus~\cite{castro1997thalamocortical, jones2001thalamic, Haber_Basal_2009}. On the other hand, cerebellum and cortex are indirectly connected through the subcortex: via the cerebello-thalamo-cortical and the cortico-ponto-cerebellar pathways~\cite{Palesi_Contralateral_2017}. Here, we found the brain to be organized into five macroscopic network modules of mixed ROIs, underlining that those extensive pathways form ordered brain-wide circuits surpassing the macro-anatomical boundaries between cortex, subcortex, and cerebellum (see Figs.~\ref{fig:Figure1}a,b and~\ref{fig:Figure2}).

The five network modules contain ROIs associated with low-level sensory and motor processing, emotion, and high-level cognition including memory, attention, executive control, language and social cognition. Superimposed upon this extensive functional diversity, each module exhibits a degree of functional bias. Community C1 largely comprises cerebellar and subcortical structures (diencephalon, brainstem and pallidum), prominently featuring regions related to visceral or autonomic signal processing, emotion and motor control. C2 comprises primarily anterior and posterior cortical regions together with subcortical structures (mostly thalamus, hippocampus, accumbens, caudate and amygdala), featuring memory-related regions. C3 and C4 are cortically dominated and show hemispheric lateralization. In line with prior reports on functional localization, the left-hemispheric community (C4) comprises more ROIs associated with language~\cite{knecht2000language, olulade2020neural} and attention~\cite{bartolomeo2019hemispheric} than the right-hemispheric community (C3), whereas the latter comprises more regions associated with social cognition~\cite{kobayakawa2018left, gupta2025relationship}. However, C5---primarily composed of medial cortical areas---exhibits the largest proportion of ROIs related to social cognition and to internally oriented cognition.

\subsection*{Composition and function of the subcortical core}

Given the need of the brain to integrate sensory information from different modalities, initially processed by specialized regions, the existence and function of multimodal ``convergence zones'' has long been debated~\cite{Baars_Book_1988, Damasio_BrainBinds_1989, Tononi_Complexity_1994, mesulam1998sensation, fuster2005cortex}. Early studies showed the presence of hubs in the cortico-cortical connectomes of macaques, cats and humans~\cite{Scannell1993, Scannell1995, Sporns_Hubs_2007, Hagmann_Core_2008, Zamora_Hubs_2010}, which are broadly connected across sensory modalities, and densely interconnected with each other, forming a hierarchical supra-module (a `rich-club'). Then, the conjecture was put forward that such a rich-club may provide the structural backbone for centralized but distributed processing~\cite{Zamora_Hubs_2010, Heuvel_HubsHuman_2011} as postulated, for example, by the Global Workspace Theory~\cite{Baars_GWT_2005, Shanahan_BroadcastNet_2010}. 

The presence of subcortical hubs has been reported by studies focusing on specific structures~\cite{Bell_Subcortical_2016, Hansen_Brainstem_2024, cambareri2024subcortical}. Our results complete these reports by identifying a rich-club at the core of the brain's global pathways which, crucially, is dominated by subcortical regions (Fig.~\ref{fig:Figure4}a,c,d). Our \emph{in-silico} analyses, based on the propagation of external stimuli throughout the brain-wide connectome, showed that subcortical hubs emerged as the most integrative nodes (Fig.~\ref{fig:Figure5}b,c) and their lesions led to significantly larger disruption of network communication than lesions of cortical or cerebellar regions (Fig.~\ref{fig:Figure5}f,g). These results are consistent with clinical evidence that patients with subcortical lesions are more likely to suffer from coma~\cite{Parvizi_Corticocentric_2009, posner2007plum, fischer2016human}, and are less likely to recover consciousness than patients with cortical damage alone~\cite{posner2007plum, lutkenhoff2015thalamic, annen2018regional}.

The central role of the subcortical hubs is remarkable considering that the subcortex only houses 0.8\% of the neurons in the brain~\cite{VanEssen_Development_2018} while the cerebellum, containing 80\% of the neurons, did not display any global hub. This disparity highlights the subcortical component of the global rich-club as a critical network bottleneck, where widespread neural signals converge onto a rather small number of neurons, whose activity is then broadcasted globally. 

A surprising observation from our analyses is that, not only is the rich-club subcortically dominated but all subcortical structures partly contribute to it (Fig.~\ref{fig:Figure4}a, middle panel). Common understanding of brain pathways used to assign, for example, a central role to the thalamus because it is the entry point for all sensory inputs to the cortex, while other subcortical structures were usually thought to deal with more specialized functions. Instead, here we have shown that the center of the brain-wide network is shared by all subcortical structures. This is possible because subcortical structures contain ROIs of diverse connectivity profiles (Fig.~\ref{fig:Figure3}e) implying that those ROIs serve different roles for the global network architecture. Within each subcortical structure we find ROIs of markedly lower (local) connectivity, likely serving more specialized functions, while other ROIs are broadly connected across the network modules and take part in the global rich-club. This internal diversity within subcortical structures raises compelling questions about their level of specialization. The presence of both locally and globally connected ROIs opens the door to clarify how individual subcortical structures may become involved in various functions, from perception to emotional and cognitive processes.

\subsection*{Limitations and Outlook}

The construction of structural connectomes based on diffusion imaging is inherently constrained by the spatial resolution and technical capabilities of the imaging procedure and subsequent preprocessing. The occurrence of false-negative and false-positive connections is a common issue when estimating structural connectivity from diffusion imaging. To mitigate this, the connectivity matrices were thresholded while preserving the original distribution of connection lengths, thus maintaining long-range fibers. Also, based on established anatomical knowledge, direct cortico-cerebellar tracts, inter-hemispheric cerebellar connections, and ipsilateral cerebellar connections to the thalamus and brainstem were excluded. Some recent studies have investigated the structural connectivity between cortex and subcortex with a notable level of detail~\cite{Benavidez_Superior_2021, alves2022subcortical, giacometti2024differential, Kumar_Thalamic_2023}. However, these works focused on selected subcortical structures. While this choice allowed high specificity of connections, they lack the systems level viewpoint of the present study, e.g., missing the internal heterogeneity of connection profiles within these structures, or the complex composition of the rich-club.

Obtaining structural connectomes with both a high-level of detail and an overarching scale is currently a major challenge for diffusion imaging. But the onset of ultra-high field imaging---scanning \emph{ex-vivo}---will soon allow the description of the human structural connectivity at an unprecedented level of resolution and completeness~\cite{plantinga2016ultra, roebroeck2018exvivo, beaujoin2018post, 
lechanoine2021wikibrainstem}. Together with the recent release of the whole-brain wiring map of the fruit fly~\cite{flywire2024whole, lin2024network}, we are about to undergo a breathtaking time to not only explore the brain's organization, but also to inquire about its evolutionary mechanisms via comparative connectomics~\cite{Suarez_Taxonomy_2022, Puxeddu_Relation_2024}. Despite the methodological precautions of this study, our findings represent an initial map and first wiring principles towards a comprehensive description of the brain's overall architecture.

\clearpage
\section*{Methods}

\subsection*{Whole-brain structural network construction}

To create the whole-brain structural connectome we employed diffusion imaging and tractography that was already preprocessed and publicly available via the Lead-DBS software package~\cite{horn2015lead}. This tractography data is based on high-quality diffusion imaging from the Human Connectome Project~\cite{van2013wu,setsompop2013pushing}. We then parcellated the data following~\citet{Ji_Mapping_2019}, which includes subcortical structures and the cerebellum.

\paragraph*{\bf Subjects and dataset.} Data of 32 participants from the Human Connectome Project (HCP) with (``MGH HCP Adult Diffusion,'' 16 females, 16 males) was collected at the Massachusetts General Hospital~\cite{van2013wu}.

\paragraph*{\bf Structural scan.} For the diffusion weighted imaging, a high-quality protocol~\cite{setsompop2013pushing} was applied (i.e. b-value of 10,000 $s/mm^2$, high-angular resolution, and a multi-slice approach). A generalized q-sampling imaging algorithm (DSI Studio; \url{http://dsi-studio.labsolver.org}) was used for the data processing. SPM 12 was used to segment and co-register the data. 200,000 fibres were examined for each participant using Gibbs' tracking approach~\cite{kreher2008gibbs} and being restricted by a co-registered white-matter mask. The fibres were standardized into MNI space via DARTEL transforms~\cite{ashburner2007fast, horn2016toward}.

\paragraph*{\bf Cortical, subcortical and cerebellar ROI assignment.} We employed the brain-wide parcellation developed by \citet{Ji_Mapping_2019}. The cortex was first parcellated into 360 regions (180 per hemisphere) according to \citet{glasser2016multi}. The subcortical and cerebellar ROI assignment was performed based on clustering of resting-state functional connectivity, see~\citet{Ji_Mapping_2019} for details. In summary, subject-specific functional connectivity (FC) matrices were calculated by correlating the individual cortical BOLD signals with Pearson correlation. By averaging across all subject matrices, a group-averaged FC matrix was constructed. Cortical resting-state functional networks were obtained by applying the Louvain algorithm to this group-averaged FC matrix~\cite{Blondel_FastUnfolding_2008}. Next, subject-wise FC matrices indicating the correlation between the 360 cortical ROIs and 31,870 subcortical and cerebellar grayordinates spanning over the entire CIFTI space were calculated and averaged to obtain a group-representative cortical -- subcortical / cerebellar FC matrix. These grayordinates were assigned to cortical functional networks with the highest correlation and merged together based on this partition. At the same time, the parcels were constrained to major subcortical structures as determined by Freesurfer, thus adhering to the general subcortical anatomy. This resulted in a partition of 233 subcortical and 125 cerebellar ROIs based on the detected cortical functional networks.

\paragraph*{\bf Filtering false-positive structural connections.} The cortex and cerebellum are contra-laterally connected via the cortico-ponto-cerebellar and the cerebello-thalamo-cortical pathways~\cite{Ramnani_CorticoCereb_2006, Palesi_Pathways_2015, Palesi_Contralateral_2017, DAngelo_Physiology_2018}. To only display direct anatomical connections, ($i$) cortico-cerebellar tracts were filtered as diffusion imaging cannot distinguish between direct and indirect pathways. This resulted in the filtering of 8.3\% of initial connections. ($ii$) Ipsilateral tracts between the cerebellum and the brainstem (pons), ($iii$) ipsilateral tracts between cerebellum and thalamus (total of 2.0\% of initial connections) where filtered out, as well as ($iv$) direct inter-hemispheric cerebellar tracts (2.6\% of initial tracts), as the cerebellar hemispheres are indirectly connected via the vermis~\cite{DAngelo_Physiology_2018}.

\paragraph*{\bf Binarization of the population-level connectivity matrix.} The individual SC matrices for the 32 participants were averaged into one population brain-wide structural connectivity network, represented as the weighted global connectivity matrix $W$ of $718 \times 718$ (ROIs). To perform graph analysis, usually $W$ is binarized by applying a hard threshold $\theta$ such that only connections with a weight larger than $\theta$ are conserved. A binary adjacency matrix $A$ is thus defined as $A_{ij} = 1$ if $W_{ij} > \theta$, and $A_{ij} =0$  if $W_{ij} < \theta$. The value of $\theta$ controls for the final number (density) of connections. This typical hard thresholding leads to a bias since tractography overestimates shorter connections. Thus, the hard thresholding favours short-range links in detriment of long-range ones~\cite{roberts2017consistency, betzel2019distance}. 

To obtain a binary adjacency matrix $A$ preserving long-range connections, we introduce the ``relative weight distance-dependent'' (RWDD) thresholding. The goal of RWDD is to conserve the original distribution of connection lengths by performing an adaptive threshold such that $\theta = \theta _d$ depends on the distance $d(i,j)$ between two ROIs. Here, $d(i,j)$ will represent the Euclidean distance between the centers of mass of two ROIs $i$ and $j$, but it could be replaced by the fiber-length of the tracts when this information is available.
First, all links (non-zero values of the population average matrix  $W_{ij}$) are sorted into $n$ bins according to the distance between ROIs. Thus, each bin contains all links of similar length $d$. The fraction of links falling in a bin (to the total number of links) is calculated. Finally, weaker links are filtered out to meet the desired target number of links in the network by applying an adaptive threshold to each bin individually, such that the fractions of links of lengths $d$ are conserved.

The RWDD thresholding was applied with a target density for $A$ of 0.2. While targeting for a low density reduces the number of false-positive connections, unconnected components could appear (i.e., ROIs with no connections to other ROIs). To correct for this, strong connections of otherwise unconnected ROIs were kept. If an ROI had the risk of becoming disconnected, links with a weight of 1.8 standard deviations above the mean were kept. This resulted in 12 additional links conserved which would initially violate the adaptive RWDD thresholding criteria.

\subsection*{Network analyses}

Given the adjacency matrix $A$ for a network of $N$ nodes and $L$ links, the \emph{density} $\rho = L \,/\, L^0$ of the network is the fraction between the number of links $L$ present and the maximum possible number of links $L^0 = \frac{1}{2} N(N-1)$. If the nodes of a network are partitioned into M groups or modules, e.g., the division of the ROIs into cortical, subcortical and cerebellar, each of $N_\alpha$ nodes and $L_\alpha$ links, the internal connection of each group is $\rho_\alpha = L_\alpha \,/\, L^0_\alpha$ where $L^0_\alpha = \frac{1}{2} N_\alpha (N_\alpha -1)$. The connection density (probability) between two groups (e.g., cortical and subcortical ROIs) is $\rho_{\alpha \beta} = L_{\alpha\beta} \,/\, L_{\alpha\beta}^0$, with $L_{\alpha\beta}^0 = N_\alpha \, N_\beta$ being the number of all possible connections between groups $\alpha$ and $\beta$.

The \emph{degree} $k_i$ of a node is the number of nodes to which node $i$ connects. The \emph{participation coefficient} $p_i$ is a measure of how a node distributes its links among modules~\cite{Guimera_RolesJSM_2005}. Here we employed the unbiased participation index defined in~\citet{Klimm_Roles_2014}, which takes into account the relative sizes of communities. In a network divided into $M$ modules, the participation vector $\mathbf{p}_i$ of the network is defined as the vector of length $M$ where $P_{i,m}$ represents the probability of a node $i$ to connect to module $m$. Then, the participation index of $i$ is:
\begin{equation}
	p_i = 1 - \frac{M}{\sqrt{M-1}} \sigma \left( \mathbf{p}_i \right) \, ,
\end{equation}
where $\sigma \left( P_i \right)$ is the standard deviation of the participation vector. When $p_i = 0$, it indicates that $i$ is only linked to other nodes within the community it belongs, whereas a value of $p_i = 1$ indicates that $i$ is equally likely connected across all modules in the network.

A \emph{rich-club} consists of supra-module of highly interconnected hubs~\cite{Zhou_RichClub_2004}. The quantitative definition of a rich-club is an undisclosed problem since (in most cases) no unique set of nodes can be strictly defined be the rich-club of a network. However, the presence or not of a rich-club can be discriminated evaluating the $k$-density $\Phi(k')$ function, which is the internal link density between the nodes with $k$ larger than a given $k'$ such that:
\begin{equation}
	\Phi(k') = \frac{L_{k'}}{N_{k'}(N_{k'} - 1)} \, ,
\end{equation}
where $N_{k'}$ is the number of nodes with degree $k > k'$ and $L_{k'}$ is the number of links between
them. Iteratively computing $\Phi(k')$ for $k' = 0, 1, 2, \ldots, k_{max}$, if the function monotonically increases, then we can assure a rich-club is present in the network. To perform a comprehensive analysis, instead of taking a single threshold, we studied the resulting rich-club organization over various thresholds in the range of $\Phi(k') = [0.8, 1.0]$. For comparative purposes, $k$-density was also estimated out of an ensemble of 100 random graphs conserving the degree distribution, using the link-switching method~\cite{Katz_Rewiring_1957}. 

\emph{Community detection}. We applied the Leiden community detection algorithm~\cite{Traag_Leidenalg_2019} to the brain-wide structural matrix. To account for different resolution parameters, we first applied the `RB' null-model~\cite{Reichardt_Partitioning_2007}. For each step of size of 0.05 in the range of $\gamma \in [0.6, 1.4]$, the algorithm was run 100 times and the partition with the highest quality value selected. Finally, the partition of this set with the highest modularity index~\cite{Newman_Modularity_2004} was chosen, see Supplementary Fig.~\ref{fig:FigureS1}.

Community detection was performed with the \emph{leidenalg} Python package, \url{https://pypi.org/project/leidenalg/}. All other network analyses were carried using the \emph{pyGAlib} Python package, \url{https://github.com/gorkazl/pyGAlib}.

\subsection*{Measuring integration and segregation}

Integration and segregation of the brain-wide network were estimated using the definitions introduced in~\citet{Zamora_Hubs_2010}. This approach accounts for the propagation of external stimuli throughout the network and the subsequent collective responses by groups of brain regions. While such propagation and collective responses could be broadly represented, for the exploratory purposes of the present study, we employed a linear dynamical approach that allows to analytically estimate the influence that one node $j$ exerts over another $i$ encompassing all possible paths, of all lengths, from $j$ to $i$.

\paragraph*{\bf Stimulus-response matrices.} Assuming the dynamics of the network (represented by its connectivity matrix $A$) are governed by a linear multivariate autoregressive process, the temporal evolution of the system is given by:
\begin{equation}  \label{eq:Autoregressive}
	\dot{\boldsymbol{x}} = - \frac{\boldsymbol{x}}{\tau} + A \, \boldsymbol{x} \, ,
\end{equation}
where $x(t) = [x_1(t), x_2(t), x_3(t), \ldots, x_N(t)]^T$ is the (column) vector of the temporal states $x_i(t)$ for all nodes, and $\tau$ is a leakage or dissipation time-constant. In this linear system, the temporal response $R_{ij}(t)$ of node $i$ at times $t > 0$ to a unit stimulus applied at node $j$, at time $t =0$, can be analytically estimated~\cite{Gilson_DynCom_2019, Zamora_Chaos_2024}. Assuming that initially all nodes receive a unit stimulus, the temporal pair-wise responses are obtained as:
\begin{equation}
	R(t) = e^{Jt} - e^{J^0 t} \, ,
\end{equation}
where $J_{ij} = -\frac{\delta_{ij}}{\tau} + A_{ij}$ is the Jacobian matrix of the linear dynamical system and $J^0_{ij} = - \frac{\delta_{ij}}{\tau}$ corresponds to the trivial leakage through a node, due to a perturbation applied on itself. This pair-wise response encompasses all network effects from $j$ to $i$ acting at different time scales along all recurrent paths of different lengths. If $\lambda_{max}$ is the largest eigenvalue of the connectivity matrix $A$, the responses will converge to zero after a transient response only if $\tau < 1 / \lambda_{max}$. Otherwise, if $\tau > 1 / \lambda_{max}$, the system diverges. For the calculations here, we considered 
$\tau = 0.5 \, ( 1 / \lambda_{max})$.

So far, $R_{ij}(t)$ represents the temporal evolution of the pair-wise responses. For the estimation of integration and segregation, the total response of node $j$ to $i$ were obtained as the accumulated response over time, by computing its integral over time (or area-under-the-curve) from $t = 0$ to $t = \infty$),
\begin{equation}
	R_{ij} =  \int_0^\infty R_{ij}(t) dt \, .
\end{equation}
In the following, we will refer to this $N \times N$ matrix $R$ as the `pair-wise response matrix'. The stimulus-response dynamics and the matrices $R$ were obtained using the \emph{SiReNetA} Python package, \url{https://github.com/mb-BCA/SiReNetA/tree/master}.

\paragraph*{\bf Integration capacity.} Here, we define integration as the sensitivity of one node (or of a set of selected nodes) to external stimuli applied to the network~\cite{Zamora_Hubs_2010}. Let $X$ be the set of all nodes in a network and $H \subset X$ a subset of arbitrarily chosen nodes. Then, the integration capacity $I(H)$ is defined as the joint response by the nodes in subset $H$, conditional to the simultaneous stimuli applied at all other nodes, the complementary set $X-H$. Formally,
\begin{equation} 
	I(H) = \mathcal{R}(H | X-H) \, ,
\end{equation}
where $\mathcal{R}$ is a generic response function. Note that the nodes in $H$ are not stimulated, instead, they ``passively listen'' to the stimuli applied everywhere except on them. This definition could be applied for an arbitrary dynamical processes running on the network. Here, we consider the linear dynamical process in Eq.~(\ref{eq:Autoregressive}). Given the vector of unit stimuli $\mathbf{s}(X-H)$ with $s_i = 0$ if $i \in H$, and $s_i = 1$ for $i \in X-H$, the temporal evolution of the response matrix is
\begin{equation}
	R(t; X-H) = \left( e^{Jt} - e^{J^0 t} \right) \, \mathbf{s}(X-H)  \, ,
\end{equation}
and the (temporally) accumulated pair-wise responses are
\begin{equation}
	R_{ij}(X-H) =  \int_0^\infty R_{ij}(t; X-H) dt \, .
\end{equation}
Finally, the integration capacity of the selected nodes $H$ is reduced to a number summing their responses:
\begin{equation}
	I(H) = \sum_{j \in H \atop i \in X-H } R_{ij}(X-H).
\end{equation}
For the results in Fig.~\ref{fig:Figure5}b, $H$ is considered as the set of 30 ROIs with largest degree $k$, and $X-H$ are the remaining 688 ROIs. The hub sets $H$ are always ranked based on their global degree, as in Fig.~\ref{fig:Figure4}a, but the selection for the different cases are based on the community the ROIs belong to, in either the anatomical or the network-based division. Hence, the colored bars all correspond to the integration capacity of 30 hubs, but only for those that belong to a specific module. The results in Fig.~\ref{fig:Figure5}c are computed equivalently but considering hub sets $H$ of increasing size, from $|H| = 1$ to $|H| = 170$, adding one hub at a time to the set $H$.

\paragraph*{\bf Segregation.} We define segregation as the loss of communication (influence) suffered between the modules of a network, after a node or a set of nodes has been lesioned~\cite{Zamora_Hubs_2010}. Let $\mathcal{P}$ be a partition of a network into $M$ modules and $C_m$ the set of nodes in module $m$. We define modular integration $I_\mathcal{P}(X)$ of the system $X$, as the amount of communication between modules. In this case, we evaluate $I_\mathcal{P}(X)$ as the sum of pairwise responses of the nodes in one module $n$, to the stimuli applied at the nodes of module $m$. Let $R_{nm}$ be the reduced modular response matrix of size $M \times M$, evaluated summing the responses of all nodes in $n$, to stimuli applied at $m$,
\begin{equation}
	R_{nm}(X) = \sum_{i \in C_n} \sum_{j \in C_m \atop  m \neq n} R_{ij} \,.
\end{equation}
Then, modular integration is calculated as the sum of the extra-diagonal entries of matrix $R_{nm}$, such that:
\begin{equation}
	I_\mathcal{P}(X) = \sum_{n=1}^M \sum_{m = 1 \atop m \neq n}^M R_{nm} \, .
\end{equation}
Assume now that a set of nodes $H \subset X$ is lesioned, i.e., deleted from the network, and modular integration is recomputed for the same partition but in the lesioned network, namely, $I_\mathcal{P}(X-H)$. Segregation is thus defined as the (normalized) loss of modular integration in the network after the lesion of set $H$, compared to the modular integration in the original `healthy' network:
\begin{equation}
	S(H) = 1 - \frac{I_\mathcal{P}(X-H)}{I_\mathcal{P}(X)}  \, .
\end{equation}
For the segregation results in Fig.~\ref{fig:Figure5}, first the reduced modular response matrix $R_{nm}(X)$ and the corresponding modular integration $I_\mathcal{P}(X)$ for the original--unlesioned--network were evaluated in both the anatomical and the network divisions. Then, targeted lesions were applied for various cases and the resulting modular integration $I_\mathcal{P}(X-H)$ was recomputed. This implied to apply a lesion to the network for the chosen set of nodes $H$ by deleting the corresponding rows and columns in the connectivity matrix $A$, and to recompute the resulting pair-wise response matrix for the lesioned connectivity, with stimulus vector $\mathbf{s}(X-H)$ where $s_i = 0$ if $i \in H$ and $s_i = 1$ for $i \in X-H$. Figure~\ref{fig:Figure5}e reports the losses in modular integration between the modules, $R_{nm}(X-H) \,/\, R_{nm}(X)$, displayed as percentages, after the 30 nodes with largest degree had been lesioned. Figure~\ref{fig:Figure5}f reports the segregation values when lesions were performed targeting either the 30 global hubs (black), or the 30 nodes with largest degree in each community (colored bars). Finally, for the results in Fig.~\ref{fig:Figure5}g were computed equivalently, but applying lesions of different sizes, from $|H| = 1$ to $|H| = 170$, by orderly including the nodes with largest degree, one-by-one, to the lesioned set $H$.

\begin{acknowledgements}
We thank Egidio d'Angelo, Claudia Gandini Wheeler-Kingshott and Fulvia Palesi for the insightful and inspiring recommendations and discussions. 
J.S. is a FI fellow of AGAUR, Generalitat de Catalunya and Fondo Social Europeo (2025 FI-2 00170). 
X.K. received support by a short-term fellowship of the European Molecular Biology Organization (Grant No. 7366).
G.D. received funding from Grant PID2022-136216NB-I00 funded by MICIU/AEI/10.13039/501100011033, by "ERDF A way of making Europe",  ERDF, EU, and by Project NEurological MEchanismS of Injury, and Sleep-like cellular dynamics (NEMESIS) (ref. 101071900) funded by the EU ERC Synergy Horizon Europe.
G.Z.L. was supported by the European Union's Horizon 2020 Framework Programme for Research and Innovation under the Specific Grant Agreement No. 945539 (Human Brain Project SGA3), and by ERDF-Project Brain dynamics, No. CZ.02.01.01/00/22-008/0004643.
\end{acknowledgements}

\section*{Author declaration}

Authors declare no conflicts of interest.

\section*{References}


\clearpage
\newpage
\widetext


\setcounter{equation}{0}
\setcounter{figure}{0}
\setcounter{table}{0}
\setcounter{page}{1}
\setcounter{section}{0}
\makeatletter

\renewcommand{\theequation}{S\arabic{equation}}
\renewcommand{\thefigure}{S\arabic{figure}}
\renewcommand{\thesection}{S\arabic{section}}
\renewcommand{\bibnumfmt}[1]{[S#1]}
\renewcommand{\citenumfont}[1]{S#1}

\begin{flushleft}
	{\Large
		Supplementary Information for: \\
		\vspace{0.5cm}
		\textbf{The global communication pathways of the human brain transcend the cortical - subcortical - cerebellar division}
	}
	\newline
	\newline

\end{flushleft}


\begin{figure*}[ht]
	\centering
	\includegraphics[width=0.9\textwidth,clip=]{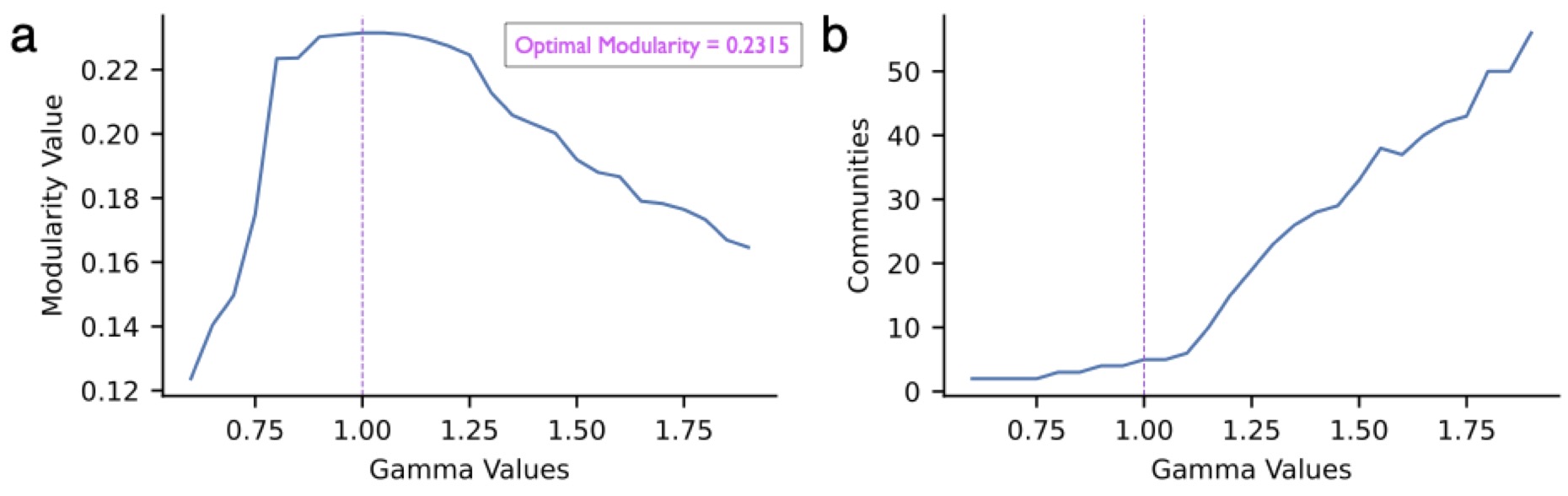}
 	\caption{		\label{fig:FigureS1}
	{\bf Community Detection, finding optimal modularity over different gamma values.}
	{\bf a}, Optimal modularity value (over 100 iterations per gamma value) for different resolution parameters gamma.
	{\bf b} The number of communities of the optimal partitioning for different resolution parameters.
	} 
\end{figure*}


\begin{figure*}[hb]
	\centering
	\includegraphics[width=0.9\textwidth,clip=]{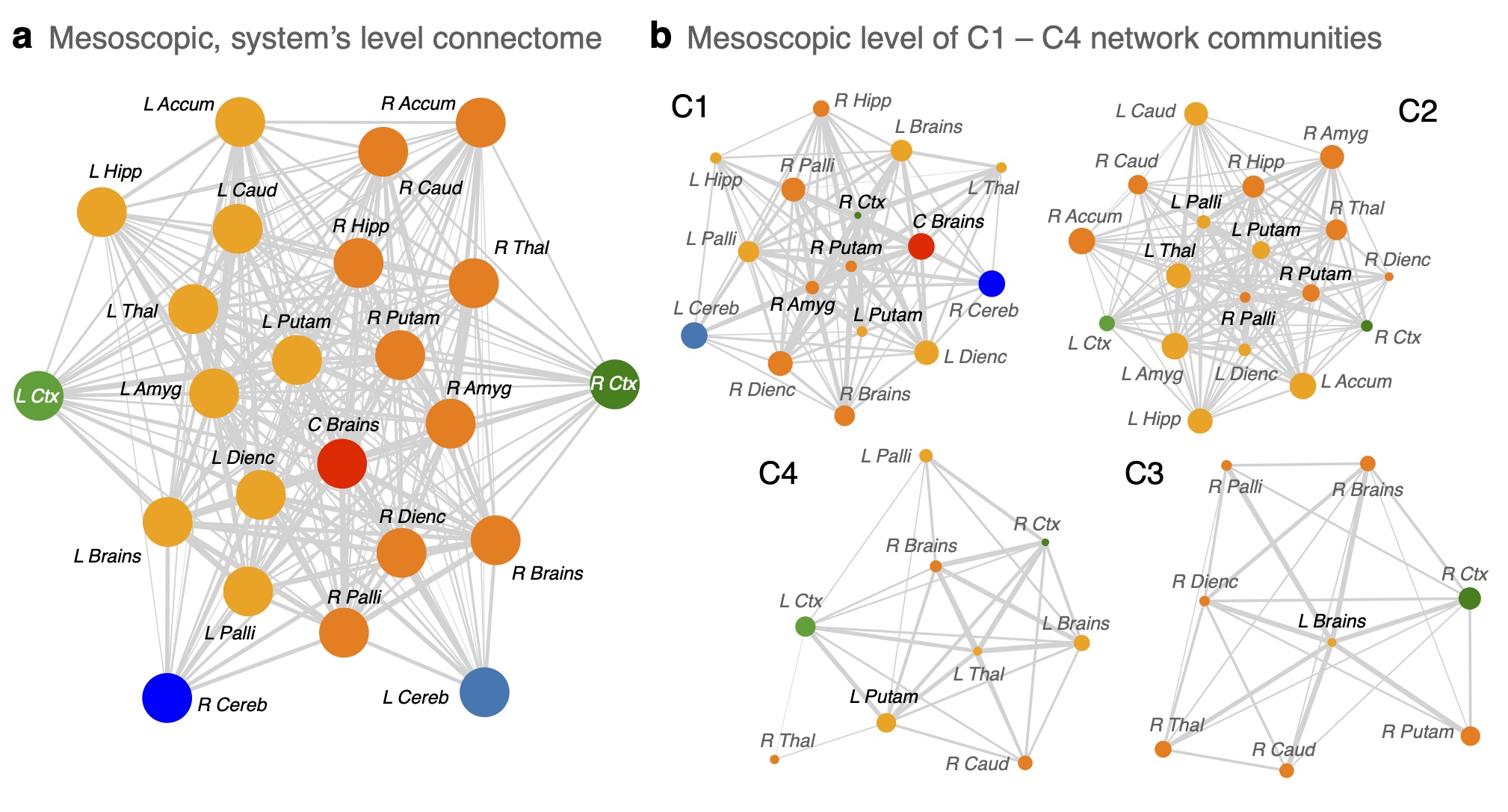}
 	\caption{		\label{fig:FigureS2}
	Visualization of the structural connectivity at the level of anatomical components comprising the cortex, the cerebellum and nine subcortical nuclei; {\bf a} for the brain-wide network and {\bf b} the subnetworks representing network communities C1 to C4. Community C5 is omitted for its simplicity as it only contains three regions. Node diameters are proportional to the fraction of ROIs that each anatomical component dedicates to the community. Hence, all diameters are equal in {\bf a}. Link widths reflect the connection probability between components, measured as the fraction of ROIs in both components that are connected, relative to all possible connections they could have (if all ROIs in both components were connected).
	} 
\end{figure*}

\end{document}